\newcommand{\eq}{\begin{equation}}
\newcommand{\en}{\end{equation}}

\documentclass[]{spie}  

\usepackage[]{graphicx}

\title{The Debris Disk Explorer: a balloon-borne coronagraph for observing debris disks}

\author{Lewis C. Roberts Jr\supit{a}, Geoffrey Bryden\supit{a}, Wesley Traub\supit{a}, Stephen Unwin\supit{a}, John Trauger\supit{a}, John Krist\supit{a}, Jack Aldrich\supit{a}, Paul Brugarolas\supit{a}, Karl Stapelfeldt\supit{b}, Mark Wyatt\supit{c}, David Stuchlik\supit{d}, James Lanzi\supit{d}
\skiplinehalf
\supit{a}Jet Propulsion Laboratory, California Institute of Technology, 4800 Oak Grove Dr.,Pasadena CA 91109, USA;\\
\supit{b}NASA Goddard Space Flight Center, Exoplanets and Stellar Astrophysics, Code 667, Greenbelt, MD, 20771, USA\\
\supit{c}Institute of Astronomy, University of Cambridge, Madingley Road, Cambridge CB3 OHA, UK\\
\supit{d}NASA Wallops Flight Facility, Wallops Island, VA 23337\\
}

\authorinfo{(Send correspondence to L.C.R.)\\L.C.R.: E-mail: lewis.c.roberts@jpl.nasa.gov, Telephone: 1 818 354 2503}

\begin{document} 

\maketitle


\begin{abstract}

The Debris Disk Explorer (DDX) is a proposed balloon-borne investigation of debris disks around nearby stars. Debris disks are analogs of the Asteroid Belt (mainly rocky) and Kuiper Belt (mainly icy) in our Solar System. DDX will measure the size, shape, brightness, and color of tens of disks. These measurements will enable us to place the Solar System in context. By imaging debris disks around nearby stars, DDX will reveal the presence of perturbing planets via their influence on disk structure, and explore the physics and history of debris disks by characterizing the size and composition of disk dust.

The DDX instrument is a 0.75-m diameter  off-axis telescope and a coronagraph carried by a stratospheric balloon. DDX will take high-resolution, multi-wavelength images of the debris disks around tens of nearby stars. Two flights are planned; an overnight test flight within the United States followed by a month-long science flight launched from New Zealand. The long flight will fully explore the set of known debris disks accessible only to DDX. It will achieve a raw contrast of $10^{-7}$, with a processed contrast of $10^{-8}$. A technology benefit of DDX is that operation in the near-space environment will raise the Technology Readiness Level of internal coronagraphs, deformable mirrors, and wavefront sensing and control, all potentially needed for a future space-based telescope for high-contrast exoplanet imaging.

\end{abstract}




\section{Science Justification}

Collisions between asteroids and sublimation of comets produce an ongoing supply of dust, which can be observed via its scattered light and thermal emission.  Such disks are very common. Extrapolating from the $\sim$25\% of Sun-like stars with debris disks detected by the Herschel Space Observatory\cite{eiroa2011} reveals that most, if not all, Sun-like stars have orbiting material\cite{greaves2010}. 

Two general characteristics of debris disks make them attractive targets for observational exploration of planet formation. First, although most of a system's mass is concentrated in large planets, dust contributes most of the surface area. Thus, the scattered light and the thermal emission of planetary systems are dominated by the debris, and debris disks are generally much easier to image than planets. Images of exoplanets in the HR 8799, and beta Pic systems, for example, were all preceded by detections of orbiting debris in those same systems.\cite{marois2008,lagrange2009}  Second, since debris disks are so common, they are found nearby. 

These general characteristics of debris disks --- nearby and bright, compared to their neighboring planets --- make them compelling targets for studying exoplanets. Local structures formed by a planet's gravitational pull can be used to pinpoint the location of individual
planets, while dust belts and large-scale disk clearings reveal the architecture of the overall planetary system. With the brighter debris tracing the location of underlying faint planets, detailed images of debris disks provide a unique opportunity to map the planetary systems of our nearest neighbors.

Unfortunately, most observations of debris disks have insufficient resolution to discern the radial dust distribution. The disks are generally first detected via their unresolved thermal emission in the far-infrared. Observations by the Spitzer Space Telescope have identified dust orbiting hundreds of nearby stars\cite{Bryden06},\cite{su06}, including Sun-like stars known to have planets \cite{Bryden09}. Herschel Space Observatory surveys have found additional fainter disks (e.g., Eiroa et al. 2011), but also have marginally resolved nearly 100 of the brightest disks. This spatial information is valuable for determining the rough location of the dust and for understanding the properties of the emitting dust. Herschel has such a long-wavelength that its resolution is comparable to the size of the disks themselves, resulting in at best marginally resolved images.

Optical light observations can achieve orders of magnitude better resolution than far-IR observations, providing an opportunity for detailed imaging of disk sub-structure. The Hubble Space Telescope (HST) has successfully observed 15 debris disks, producing a set of iconic images such as the famous ring around Fomalhaut. HST has exhausted the list of nearby bright disks that meet its performance requirements; a new optical-light facility optimized for debris disk science is needed to expand on these ground-breaking discoveries.
 
\begin{figure}
   \begin{center}
   \begin{tabular}{c}
   \includegraphics[height=6cm]{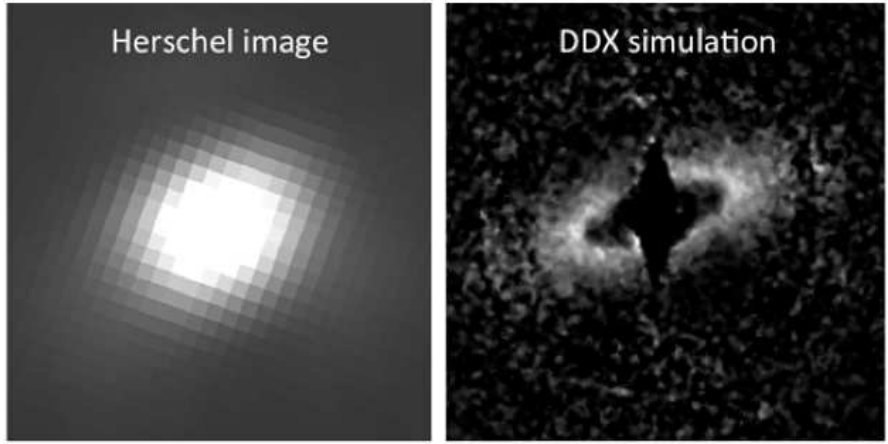}
   \end{tabular}
   \end{center}
   \caption[] 
    { \label{herschel_comparison} The Debris Disk Explorer (DDX) will make discoveries not possible with any existing or planned telescope. 49 Ceti, a debris disk that HST failed to detect, will be easily imaged by DDX at a resolution 25 times better than Herschel, revealing underlying disk structure such as planetary gaps and planet-induced eccentric rings. The image has a field of view of 7.5 arcseconds.}
\end{figure}

The Debris Disk Explorer (DDX) is a proposed  mission to observe planetary system structures around nearby stars with a telescope on a gondola carried by a stratospheric balloon. The end product will be a set of high-resolution images over four optical-wavelength bands for each targeted debris disk. An ultra-long duration balloon (ULDB) flight from New Zealand will image tens of disks with a combination of contrast and angular scale that cannot be achieved by any other existing or upcoming facility (e.g., HST, JWST, or ground-based adaptive optics coronagraphs); see Figure \ref{herschel_comparison}. The resulting images will be used to detect signatures of long-period planets, to observe planetary system architectures, and to characterize the dust properties, thereby improving our understanding of how planetary systems form and evolve.

\subsection{Planetary System Architectures}

Debris disk images can be used to probe the overall architecture of planetary systems. In the Solar System, for example, the terrestrial and gas-giant planet regions are bounded by the asteroid and Kuiper belts. By observing just the dusty belts, a distant observer could infer the location of the intermediate planets. Similarly, the four planets of HR 8799\cite{marois10} could have been predicted based on their location in a gap between interior and exterior belts of dust\cite{su09}.

High resolution images are essential because of ambiguity in the dust location based solely on the spectral energy distributions of an unresolved debris disks. Marginally resolved images with Herschel provide a rough estimate for the dust location, but are unable to distinguish between radially narrow or broad distributions, or to determine if the emission is in multiple narrowly confined belts\cite{churcher11,wyatt12}.

Our first key science objective is to resolve the overall architectures of a sample of nearby debris disks. In particular, our aim is to distinguish between broadly distributed emission and dust that is confined to narrow rings. By making this distinction, we will learn about the status of planet formation within each observed system. A sharply confined belt of material implies that the inner region has been cleared out by unseen planets\cite{quillen06,mustill12}. A broadly distributed disk, on the other hand, is suggestive of failed planet formation. Such a disk indicates that the building blocks of planet formation (planetesimals) still exist today throughout the disk and were never able to accumulate into a larger planet, nor were they cleared out by such a planet.

By resolving planetary system architectures, we can compare the locations and widths of their asteroid and Kuiper belts with those of the Solar System and determine whether or not the Solar System's planetary system is normal or is an outlier. While many planet formation theories predict that configurations similar to the Solar System's --- with large gaseous planets forming just outside the snow line and rocky planets interior to it --- should be common, there are undoubtedly many alternate paths for planetary systems to follow. Indeed, the planets that have been detected around other stars are much different than the planets of the Solar System. They range from hot Jupiters with periods of a few days to young massive planets far beyond Pluto's orbit.  The orbits are often highly eccentric and sometimes retrograde.

\subsection{Signatures of Long-Period Planets}

While many planets have now been identified around other stars, the most successful detection techniques (radial-velocity measurements and transit detections) are limited to orbits close to the parent star. At planet-star separations $>10$ AU, direct imaging is more effective, but thus far has been limited to detections around a few stars that are brighter and younger than the Sun

Of greatest importance for DDX, the structure of the disks can be used to predict the planets’ locations. The warp of beta Pic’s disk, for example, was predicted to be driven by an unseen companion\cite{mouillet97}; the orbit of the later detected planet is consistent with the prediction (Lagrange et al. 2010). 

Just as the discovery of short-period planets sparked a wave of new theories of planet formation, the presence of planets far from their central stars (at 10’s to over 100 AU) provides an important new constraint to guide our understanding of how planetary systems form and evolve. The standard core-accretion theory of planet formation cannot produce giant cores outside of a few 10's of AU and while there can be scattering outward (e.g., the``Nice'' model\cite{tsiganis05}), such scattering has not been shown to produce a stable configuration like HR 8799's\cite{sally09}. Direct gravitational collapse, meanwhile, should also produce a population of higher-mass brown dwarfs, which have not been observed\cite{kratter10}. How gas giant planets might form in or migrate to the outer reaches of planetary systems remains an unanswered question.

\subsection{Debris Disk Physics}

Our understanding of the fundamental physical processes operating within debris disks is limited, resulting in ambiguous interpretation of their history and evolution. The observed dust in any specific system might originate from a steady collisional cascade of small asteroids, from large single events like the Moon-forming impact, from system-wide instability like the Late Heavy Bombardment, or from swarms of sublimating comets.  Measuring the dust properties (size and composition) is crucial for distinguishing between these possible origins, and so for understanding the underlying physics of debris disks and thereby determining how they form and evolve.

Albedo (the fraction of incident light that is scattered, rather than absorbed) is a fundamental measurable that provides a valuable constraint for determining the dust grain size and composition. Albedos derived from HST-resolved disks range from darker than asphalt (5\%) to brighter than fresh snow (89\%). High albedos generally require freshly-processed icy composition, while lower albedos imply rocky dust. The makeup of the dust, whether rocky or icy, indicates if the parent bodies are asteroidal or cometary.

Besides albedo, DDX will also measure disk color. The AU Mic system serves as a example of what we can learn from DDX color imaging. While most disks observed by HST are red, AU Mic's dust is unusually blue, indicating grains that are small compared to the observing wavelengths (500--900 nm). Furthermore, the disk gets bluer outward, suggesting that the grains are smaller in the outer disk. These observations provide important information about how the dust is removed from the disk.


\section{Mission Overview}

The Debris Disk Explorer (DDX) will observe debris disks around nearby stars with a 0.75-m diameter telescope on a gondola carried by a stratospheric balloon.  The challenge for imaging at optical wavelengths is to block out the light from the central star, revealing the surrounding dust. Though brighter overall than their neighboring planets, debris disks are tenuous, only scattering a small fraction of the star's light. High-contrast imaging at contrast ratios $<10^{-7}$ is needed. To achieve this contrast ratio requires a high-performance coronagraph that is maintained in a stable, i.e., space-like environment. Balloon-borne stratospheric observations provide the necessary thermal and atmospheric stability at a fraction of the cost of a space mission.

\begin{figure}[thb]
   \begin{center}
   \begin{tabular}{c}
   \includegraphics[height=9cm]{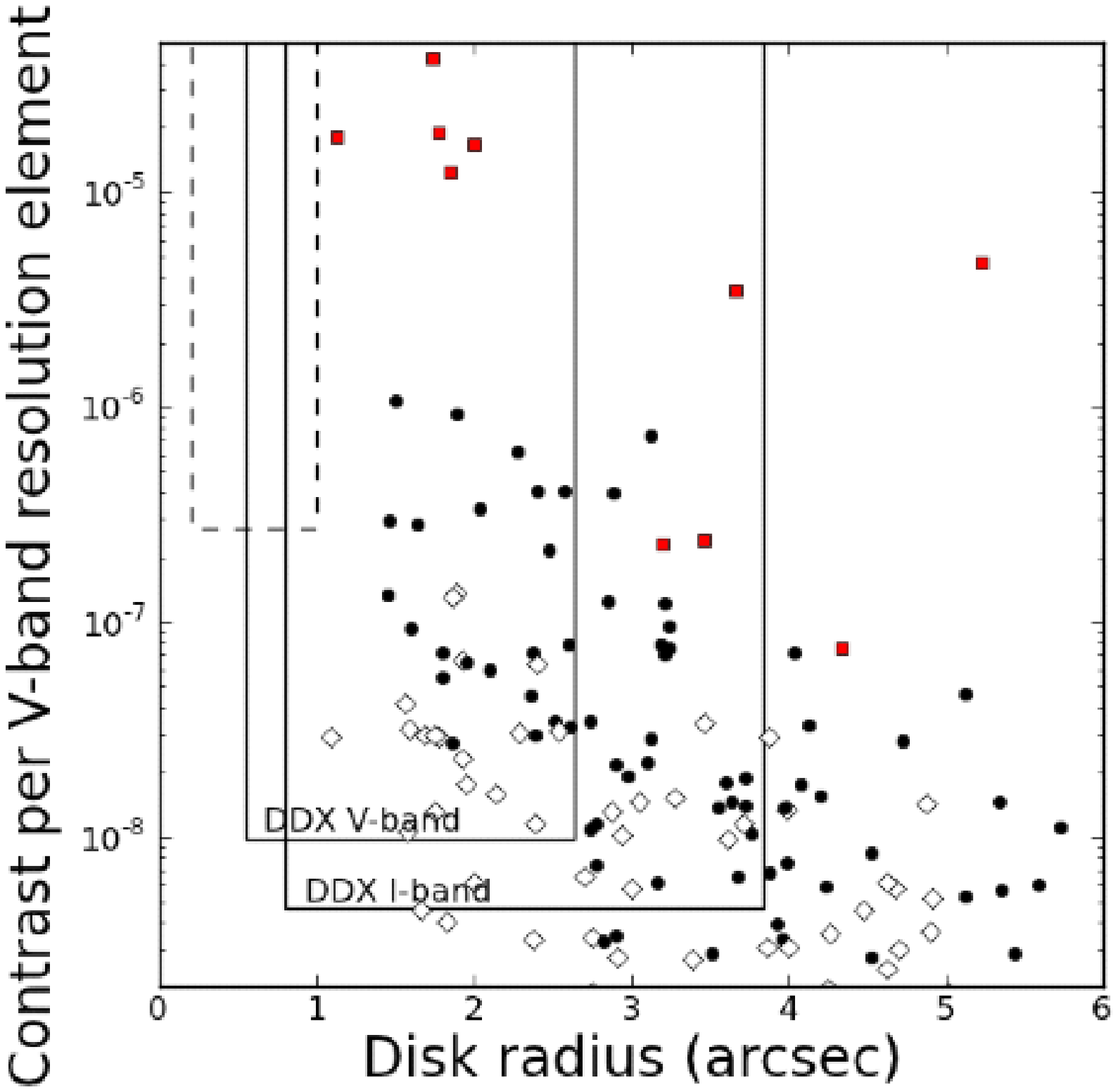}
   \end{tabular}
   \end{center}
   \caption[] 
    { \label{contrast_performance} The DDX telescope and high performance coronagraph are well matched to the observed sizes and brightness  contrasts for known debris disks. The disk radii shown here are measurements from HST images (red squares) or Herschel images (black circles), or are model estimates (white diamonds). Unlike the next generation of ground-based coronagraphs (dashed line), DDX’s science bands (solid U-shaped lines) can image many of the known disks.}
\end{figure}

Figure \ref{contrast_performance} shows the science leap forward enabled by DDX based on its contrast performance and field of view. In this figure, contrasts relative to the central star are calculated based on the Herschel observed thermal brightness of the disk (L$_{I\!R}$/L$_{*}$) and angular size of the disk. The key uncertainties in the calculation are the quantities that DDX will measure, the visual brightness (albedo) and the radial extent ($\Delta$R/R). For the predictions in Figure \ref{contrast_performance}, we assume values for albedo and $\Delta$R/R based on the range observed by HST. The disk albedo is set to 0.1, a conservative value given the observed range from 0.05 to 0.89; the ring width $\Delta$R/R is set to 0.3, consistent with the observed range.

Figure \ref{contrast_performance} also illustrates the relative strength of a modest size balloon-borne telescope for imaging disks. DDX's contrast per resolution element is a factor of ten better than the projected performance of ground-based telescopes ($10^{-8}$ vs. $\approx10^{-7}$ for GPI\cite{mcbride2011} and SPHERE\cite{kasper2012}), expressed in terms of contrast ratio per resolution element, i.e. the contrast for detecting a point source. The higher resolution of larger ground-based telescopes effectively spreads the disk over more resolution elements, making a disk of a given surface brightness harder to detect. This can be seen in Figure \ref{contrast_performance}, where the sensitivity curves have all been calculated after re-binning the data to DDX's V-band resolution, such that the ground-based sensitivity (dashed line) is somewhat worse than $10^{-7}$.

The primary mirror for the DDX telescope will be 0.75-m in diameter - large enough to meet our resolution requirement of 0.25 arcsec at optical wavelengths. The light will be split into four wavebands covering the wavelength range 500--900 nm, with two broad bands centered at 550 and 800 nm acting as our primary science bands (sufficient to measure a disk color) and two narrower bands for identifying and correcting wavefront aberrations. The expected contrast for the coronagraph after post-processing is $1\times10^{-8}$, a factor of three better than required.

\subsection{High Altitude Turbulence}

DDX operates in the stratosphere in order to avoid the blurring effects of tropospheric turbulence  which limits ground based coronagraphs, even those with adaptive optics to contrast values of about $10^{-6}$\cite{hinkley09}. For a stratosphere-based telescope at 35 km, the overhead atmosphere is reduced by a factor of 170, and, in the angular range of DDX (0.6 to 3.8 arcsec), the scattered light from the star is dramatically reduced

In addition to the free-air contribution to background contrast, we investigated the effect of local turbulence, from the gondola itself, using a laser interferometer on a balloon gondola\cite{traub08}. No evidence of any disturbance at the limit of our accuracy ($1/32$ wavelength) was found. Overall, assuming that the free-air wavefront error dominates any error from the air very close to the gondola, as is the case with a
well-designed ground-based telescope, we find that the net disturbance to the wavefront along the total path is expected to yield a speckle contrast value well below our control limit of 10$^{-8}$, so the atmospheric contribution to the speckle background is essentially absent.

\subsection{Balloon Flights}
 
After integration in the laboratory, the DDX payload will be tested in an overnight Integration and Test flight launched from Ft.~Sumner, New Mexico.  This flight will observe a handful of debris disks and point spread function calibrator stars.  These observations will be used to validate system performance and to test the various hardware components.  After the test flight, the system will be recovered.  The system telemetry and science results will be analyzed to determine the level of system performance.    If any systems are not performing according to requirements, they will be modified before the science flight, which is scheduled 10 months later.  The DDX mission's science is accomplished on a 30-night ultra-long duration balloon flight from New Zealand to South America. The 30-night flight plan is that the balloon launches toward the west, passes over South America, continues around the world, and returns to South America to land.


\section{Instrument}

\begin{figure}[hbt]
   \begin{center}
   \begin{tabular}{c}
   \includegraphics[height=8cm]{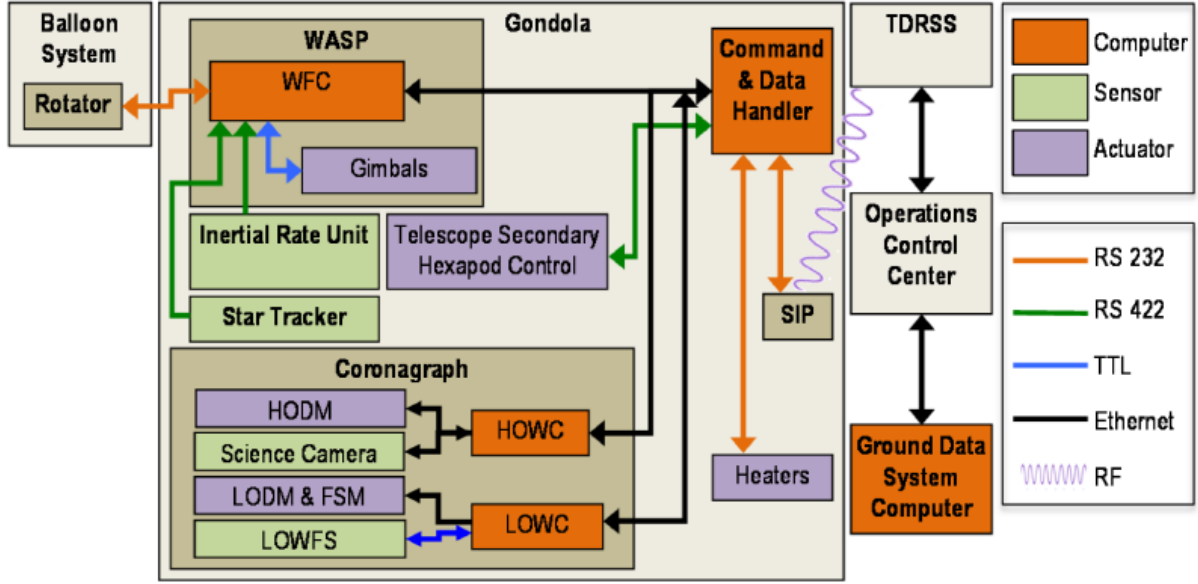}
   \end{tabular}
   \end{center}
   \caption[] 
    { \label{block_diagram} Block diagram showing the connections between the DDX subsystems. The type
of connection is indicated by the line color. The type of subsystem is indicated by block color.}
\end{figure}

The DDX Payload consists of a number of subsystems. Fig. \ref{block_diagram} shows the system block diagram. The heart of DDX is the coronagraph that enables the high-contrast images.  It is mounted to an off-axis telescope pointed by the Wallops ArcSecond Pointing (WASP) system. The power, communications and the flight computer are indicated in Figure \ref{block_diagram}. A gondola encloses all the subsystems and protects them during launching and landing.  

\subsection{Coronagraph}

The coronagraph is mounted on bipods immediately behind the PM. An optical diagram is shown in Figure \ref{optical_design}. The light encounters the following components:

\begin{figure}[bt]
   \begin{center}
   \begin{tabular}{c}
   \includegraphics[height=9cm]{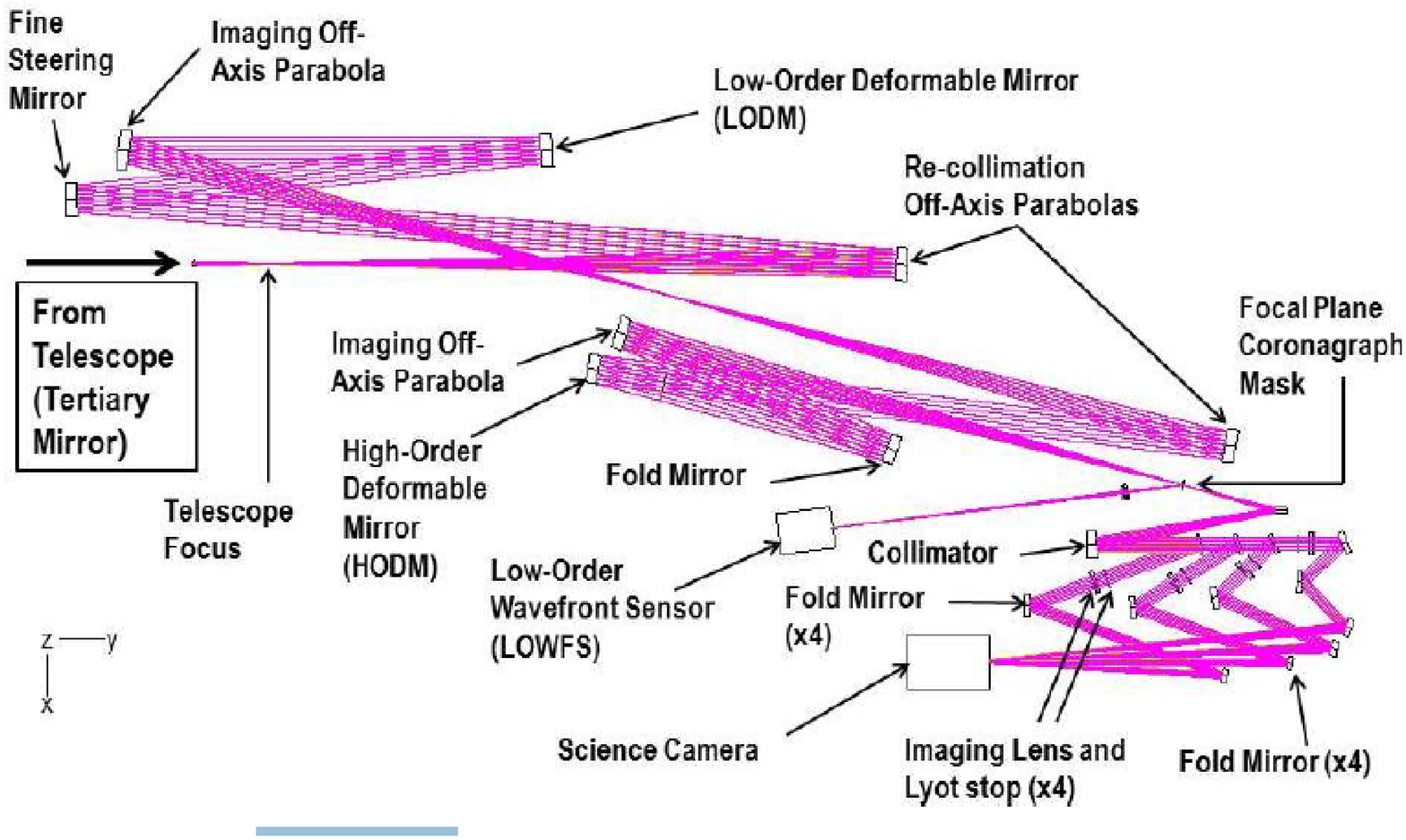}
   \end{tabular}
   \end{center}
   \caption[] 
    { \label{optical_design} The optical layout of the coronagraph instrument. Light enters at the telescope fold mirror just before the telescope focus on the left. The LOWFS uses light reflected from the focal plane mask. After the mask, the beam is split into four wavelength bands that are focused in the four quadrants of the science camera.
\\
}

\end{figure}

\textbf{Fine-Steering Mirror (FSM):} After the telescope focus, the beam is re-collimated, forming an image of the PM at the FSM, which corrects for image jitter.  The FSM is driven by feedback from the low-order wavefront sensor (LOWFS).

\textbf{Low-Order Deformable Mirror (LODM):}  After the FSM is the LODM, which corrects low-order aberrations (focus, coma, spherical, astigmatism). The LODM is a surface-normal deformable mirror from Xinetics with a 10$\times$10 array of actuators. The actuators can deflect the surface of the mirror by $\pm$2 $\mu$m (also called stroke). After the LODM, the pupil is relayed to the high-order deformable mirror.

\textbf{High-Order Deformable Mirror (HODM):} The HODM, also from Xinetics, has 48$\times$48 elements across a continuous face sheet, with a stroke of $\pm$1 $\mu$m. The HODM corrects errors on a spatial scale that is finer than the LODM. A fold mirror and off-axis parabola relay the star to the coronagraph mask.

\textbf{Coronagraph Mask:} The DDX coronagraph mask is a hybrid band-limited mask\cite{moody08}. The same basic design has yielded a
contrast of $6\times10^{-10}$ in the lab, with a stability of $0.1\times10^{−10}$ RMS over 5 hours\cite{trauger07}. The central portion of the occulting mask is reflecting, to feed the LOWFS.

\textbf{Low-Order Wavefront Sensor (LOWFS):} The LOWFS uses star-light from the coronagraph mask to measure tip/tilt and low-order wavefront aberrations using a Shack-Hartmann wavefront sensor.

\textbf{Dichroics:} Light that is transmitted past the coronagraph mask (mostly debris disk light with very little starlight) next sees a series of dichroics, which divide the beam into four wavebands.  

\textbf{Lyot Stops:} After the dichroics, each beam encounters a Lyot mask optimized for the wavelength of the band.  The masks block diffracted star-light.

\begin{figure}[th]
   \begin{center}
   \begin{tabular}{c}
   \includegraphics[height=9cm]{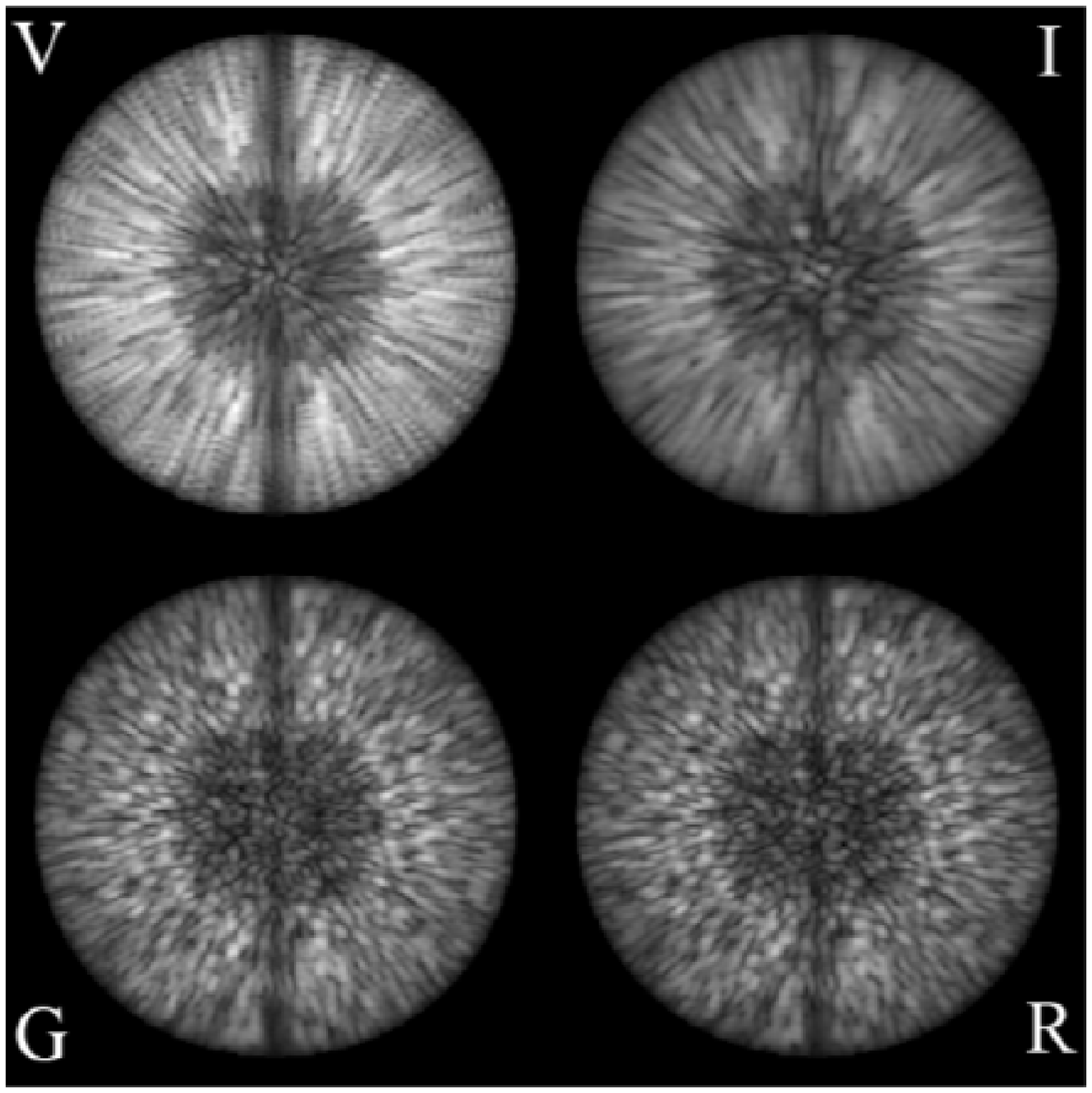}
   \end{tabular}
   \end{center}
   \caption[] 
    { \label{simulated_image} Simulated DDX field showing four channels imaged on the CCD. Channels VGRI are defined in Table \ref{filters}.  Since the V and I images have wider bandwidths, the speckles are more smeared out.  Spatial scale varies with bandpass. Each dark vertical line is a shadow of the linear field mask. The dark hole occupies the central region of each field.}
\end{figure}

\textbf{Science Camera:} All four channels are then imaged onto the same CCD, one channel in each quadrant. The science camera is a high-quality low-dark-current CCD camera. A simulated image from the camera is shown in Figure \ref{simulated_image}.

\begin{table}[bth]
  \caption{ The definition of the DDX spectral wavebands.  They provide sufficient field of view (IWA to OWA) and spatial resolution (FWHM) to image disks. } 
  \label{filters}
  \begin{center}       
  \begin{tabular}{|c|c|c|c|c|c|c|} 
  \hline
  \rule[-1ex]{0pt}{3.5ex} Band & Primary Use & Central    & Bandwidth & Lyot & IWA-OWA & FWHM \\
  \rule[-1ex]{0pt}{3.5ex} Name &   & Wavelength  & ($\Delta\lambda/\lambda)$ & Stop Size & (arcsec) & (arcsec) \\
  \hline
  \rule[-1ex]{0pt}{3.5ex}  V & Science & 550 nm   & 20\% & 67\% & 0.6-2.6 & 0.23 \\
  \rule[-1ex]{0pt}{3.5ex}  G & Wavefront sensing & 630 nm & 5\% & 65\% & 0.7-3.3 & 0.29 \\
  \rule[-1ex]{0pt}{3.5ex}  R & Wavefront sensing & 700 nm& 5\% & 61\% & 0.8-3.6 & 0.37 \\
  \rule[-1ex]{0pt}{3.5ex}  I & Science & 800 nm    & 20\% & 52\% & 1.0-3.8 & 0.46 \\
  \hline 
\end{tabular}
\end{center}
\end{table}

\subsection{Wavefront Sensing and Control}
 
Low-order modes are detected with the low-order wavefront sensor (LOWFS), which splits the input pupil into a few sub-apertures across the diameter and measures the slopes across each one. This is done by placing a lens behind each sub-aperture, and measuring the spot motion from each one and comparing it to a nominal spot location for a perfect system. The light used for this sensor is reflected from the focal-plane mask, and this acts as a spatial filter. Therefore modes greater than $\sim4 \lambda/D$ are not sensed because they are not present. Our analysis of the LOWFS gives a tip-tilt measurement limit of 0.0027 arcsec per sample, and a low-order sensing limit of 0.1 nm per sample at 6th magnitude and 500 Hz. The pointing information from the low-order sensor is used as feedback for the guidance system, which controls the FSM. Focus information is fed to the SM which is translated via its hexapod. Other low-order aberrations are corrected with the LODM.

The HODM initially flattens the wavefront using a pre-determined pattern derived from phase retrieval measurements taken during ground tests. To further reduce the speckle intensity, on-sky star observations are used to derive the wavefront, based on the residual speckles in the two 5\% filters on the science camera. During this wavefront sensing procedure, a series of four probe patterns is placed on the HODM, forming a sequence of intensity measurements that can be inverted to derive the electric field at the image plane\cite{borde06,giveon11}. This is used to determine the HODM settings that will minimize the speckles inside the dark hole region. For a V = 0 star, each iteration will take 2.5 min., and we estimate 10 iterations will be necessary, for a total of 25 min.

\subsection{Telescope} 		 
To meet our imaging requirements, the DDX telescope must have: (1) collecting area large enough to image debris discs in reasonable
times, (2) angular resolution high enough to make sharp images of debris discs, (3) wavefront that is smooth and stable enough that
speckles are either faint or removable in post processing, and (4) mass small enough to fly as a balloon payload.

Criteria (1) and (2) are met by having a pupil aperture of 0.75 m, which provides sufficient collecting area to make images of debris disks on hour-long timescales  and sufficient resolution to resolve disk structures induced by orbiting planets. The need for sharp imaging and a simple coroangraph design also requires the absence of an obstructing secondary and spider, so an off-axis, unobscured telescope is baselined. Criterion (3) demands that the telescope mirrors and structure remain stable at the sub-micron level during pointing or temperature changes, and that further correction at the sub-nm level is achieved with a wavefront sensing and control system. DDX stabilizes the telescope using stiff, low-coefficient of thermal expansion (CTE) Zerodur mirrors, a closed-loop temperature control system for the PM, and a stiff, low CTE optical support structure of composite materials. Lastly, criterion (4) is met by light weighting the primary mirror by ~90\%, and by using lightweight materials for the support structure.

\begin{figure}
   \begin{center}
   \begin{tabular}{c}
   \includegraphics[height=10cm]{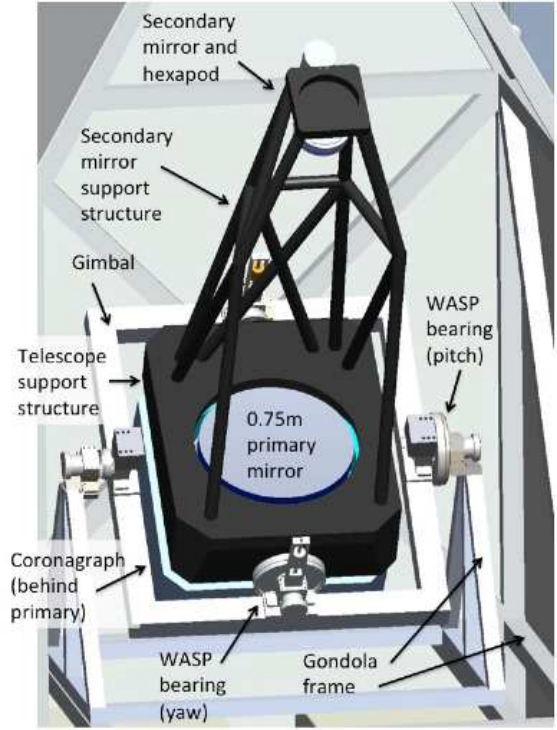}
   \end{tabular}
   \end{center}
   \caption[] 
    { \label{telescope} The off-axis design of the DDX telescope allows incoming light to reach the primary mirror (PM) with no obscuration. After reflection off the secondary mirror, light enters the coronagraph mounted behind the primary mirror.}
\end{figure}

This telescope is an off-axis Gregorian with a 0.75-m diameter, light-weighted Zerodur primary mirror, and a 0.12-m diameter secondary mirror, both coated with enhanced aluminum. The secondary mirror is mounted in a hexapod that controls its position in translation, tilt and rotation (six degrees of freedom). Fig. \ref{telescope} shows the telescope as well as the pointing system.

\subsection{Pointing System}

Pointing is a key enabling technology for DDX.  The DDX coronagraph requires the target star to be accurately centered on the coronagraph mask, to avoid leakage of starlight to the image plane. This is achieved by the successive actions of the rotator, the WASP pitch-yaw gimbals, and the FSM. The signals driving these actions are from the inertial navigation system, star tracker, LOWFS, and science camera.

The DDX coronagraph also requires that the wavefront be smooth, to avoid speckles in the image plane. This is achieved by minimizing ``beam walk'', differential motion of starlight from the primary mirror across the secondary and tertiary mirrors, and by using a low-order deformable mirror followed by a high-order deformable mirror. The signals driving these actions are from the star tracker, the LOWFS, and the science camera. The pointing requirements, capabilities, and margins are listed in Table \ref{pointing_requirements}.

\begin{table}[h]
  \caption{ Pointing stability and accuracy over 300 s (1$\sigma$ RMS per axis). } 
  \label{pointing_requirements}
  \begin{center}       
  \begin{tabular}{|l|c|c|c|} 
  \hline
  \rule[-1ex]{0pt}{3.5ex} Pointing Stage & Requirement & Capability & Margin \\
  \hline
  \rule[-1ex]{0pt}{3.5ex}  Rotator & 5 deg. & 2 deg. &150\% \\
  \rule[-1ex]{0pt}{3.5ex}  Telescope - Stability & 1.0 arcsec & 0.25 arcsec & 300\%   \\
  \rule[-1ex]{0pt}{3.5ex}  Telescope - Accuracy & 0.5 arcsec & 0.1 arcsec & 400\% \\
  \rule[-1ex]{0pt}{3.5ex}  FSM - Stability & 0.04 arcsec & 0.02 arcsec & 100\%   \\
  \rule[-1ex]{0pt}{3.5ex}  FSM - Accuracy & 0.02 arcsec & 0.01 arcsec & 100\%   \\
  \hline 
\end{tabular}
\end{center}
\end{table} 

\begin{figure}
   \begin{center}
   \begin{tabular}{c}
   \includegraphics[height=10cm]{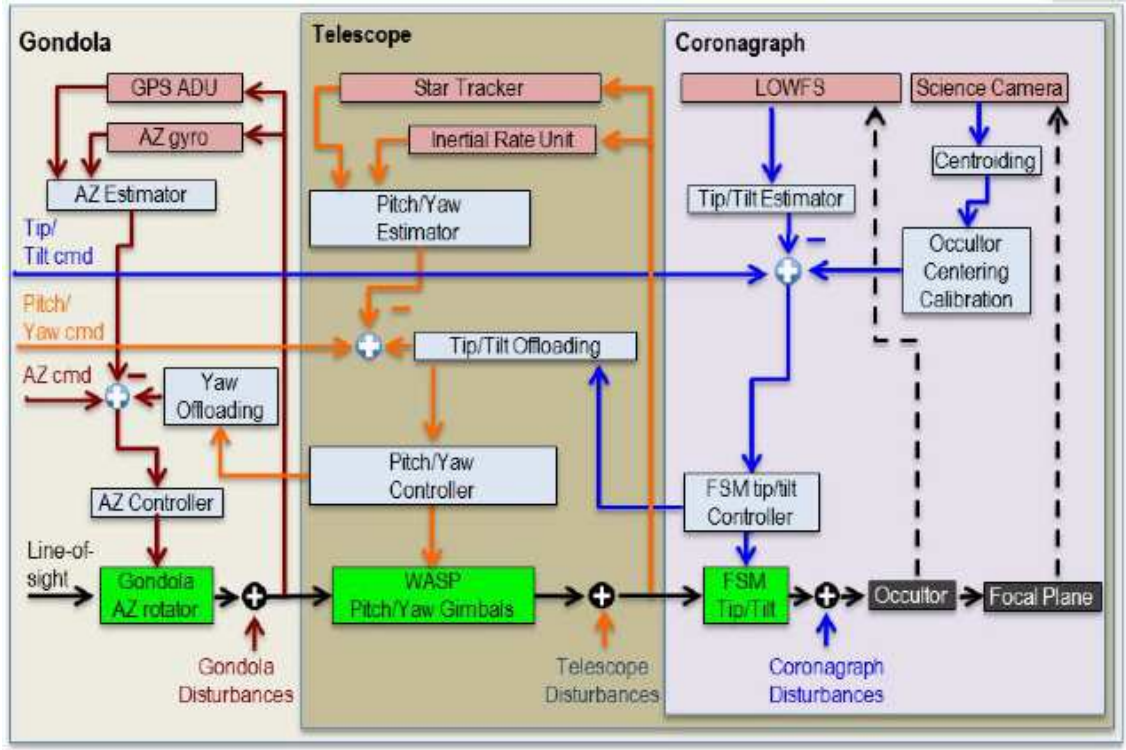}
   \end{tabular}
   \end{center}
   \caption[] 
    { \label{pointing_architecture} Pointing architecture for DDX. There are three stages of line-of-sight control (black signal):
gondola AZ rotator (red signal), WASP pitch and yaw gimbals (orange signal) and
coronagraph tip/tilt mirror (blue signal). Actuators are depicted in green, sensors in red,
algorithms in blue, and the coronagraph occulter and focal plane in dark gray.}
\end{figure}

The architecture of the real-time on-board DDX pointing system is shown in Figure \ref{pointing_architecture}. The  rotator uses its integrated single axis azimuth gyro and GPS Attitude Determination Unit, at low bandwidth ($\sim$0.1 Hz) to maintain the coarse orientation of the gondola. The Wallops Arc-Second Pointing (WASP) system moves the telescope in pitch and yaw using motors and continuously rotating bearings to avoid stiction. It is driven by the star tracker and the inertial rate unit signals  at medium bandwidth ($\sim$10 Hz) to maintain accuracy of the telescope pointing. A balloon-borne demonstration of the WASP pointing system, including the rotator, star tracker, inertial rate unit, and pitch-yaw gimbals, all of the same type as proposed for DDX, demonstrated inertial rate unit based pointing stabilities (1$\sigma$) of 0.23 arcsec in a 2011 flight, and 0.33 arcsec in a 2012 flight

The FSM  uses the LOWFS to measure the tip-tilt of the star image at high bandwidth ($\sim$100 Hz). The tip/tilt stage has a range of $\pm$33 arcsec on the sky. DDX uses a copy of the FSM flown on the PICTURE sounding rocket, which achieved an accuracy of 0.005 arcsec\cite{mendillo12}.

The pointing controller is a nested set of integral control laws. The tip/tilt signal from the LOWFS goes first to the FSM, using an integrator term and an AO control matrix for the low-order Zernike terms. In the next tier, any slow offset of the FSM is offloaded to the WASP pitch and yaw gimbals, and these, in turn, are offloaded to the rotator. The LOWFS target location for the star is acquired and then calibrated through a slow loop (0.01 Hz) using any offset seen in the science image during the previous science exposure, suppressing non-common path errors. All the pointing estimation and control algorithms run real time on the on-board computers and do not require ground post-processing.

 The main pointing disturbances are balloon pendulations, pitch-yaw bearing noise, and the stiffness of cables passing over the pitch-yaw gimbals. Balloon pendulations generate low amplitude ($<1^\circ$) low frequency ($<1$ Hz) pointing errors\cite{traub86}. These errors are rejected by the gondola and telescope pointing control systems.

\begin{figure}
   \begin{center}
   \begin{tabular}{c}
   \includegraphics[height=12cm]{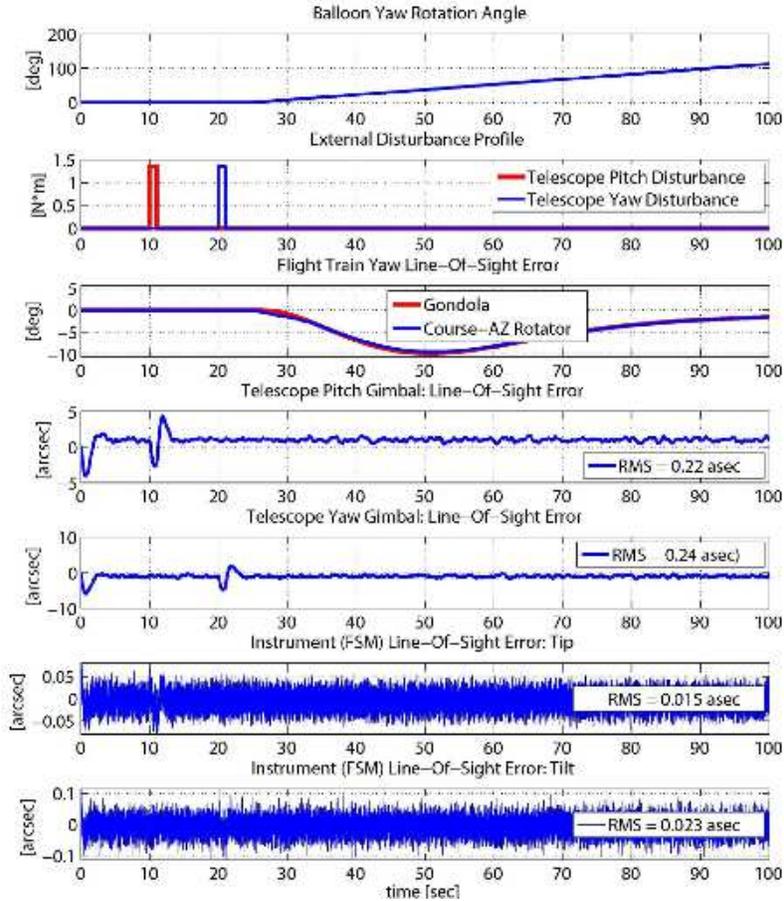}
   \end{tabular}
   \end{center}
   \caption[] 
    { \label{pointing_simulations} Results of the end-to-end simulations of the pointing system, showing that DDX is able to achieve the required pointing accuracy.}
\end{figure}

To verify that the pointing system would meet requirements, we developed an end-to-end pointing simulation\cite{aldrich2013}. The simulation includes a multi-element flight train, rotator, WASP pitch and yaw gimbals, telescope and coronagraph (with the expected DDX inertia), and FSM. Figure \ref{pointing_simulations} shows the end-to-end performance as the gondola is subject to: i. a system start up transient at time 0 s; ii. a 0.14 m-kg pitch impulse disturbance response at time 10 s; iii. a 0.14 m-kg yaw impulse disturbance response at time 20 s; iv. a balloon spin rotation starting at time 25 s. The plots show that the system handles the disturbances gracefully and that the steady state accuracies meet the allocations. This simulation gives confidence in the pointing design and corroborates the performance estimates given in Table \ref{pointing_requirements}.

\subsection{Gondola} 	
	 
The DDX gondola is constructed of aluminum tubing, welded at the joints, with bolted cross braces for easy access to the instruments. The science instrument is mounted to the gondola via a triangular yoke. A lower deck supports electronics, batteries, and telemetry. The gondola hangs from a standard NASA rotator via steel cables and a spreader bar. The rotator has a built-in sun tracker, allowing the gondola to orient its solar-cell-covered backside toward the sun during day, or its front side to track stars at night.

\subsection{Power}	
		 
DDX power is derived from lithium ion batteries  charged from solar panels mounted on the back of the gondola.  During the day the solar panels are kept facing the sun by the rotator and its embedded Sun sensors.  The solar panels on the back of the gondola
are mounted on stand-offs, so air can circulate between the panels and the gondola insulation; this heat source is modeled to dissipate quickly at sunset, leaving the instrument temperatures within tolerance.

\subsection{Telecommunications}	 

DDX uses the NASA standard Support Instrumentation Package (SIP) for our science flight. The SIP handles slow-rate data transmission to the ground via an omni-directional TDRSS antenna (with an Iridium antenna backup), and a pointed TDRSS high-gain antenna for high rate
data.  

For the entire 14-hour night, during each 5 min exposure, we download the dark hole  (the area with the highest contrast) of the previous observation V-band sub-image (799 kb/dark hole) via the omni-directional TDRSS, along with 0.8 kbps of selected housekeeping data. The housekeeping downlink data rate continues during the day. During the day, we use the TDRSS high-gain antenna (which must be steered and is controlled by the SIP) to download the entire 512$\times$512 science data subframe for all the exposures (672 Mb), the HODM (5.9 Mb) \& LODM (262.5 kb) commands and the housekeeping data for the night (384.5 Mb). This takes just under four hours to download with an assumed overhead of 20\%.




\section{Science Analysis}

A science dataset consists of at least of an hour or more of integration on a science target, along with observations of a reference star to a similar exposure depth. The non-equatorial telescope mount of DDX will cause the sky to rotate on the detector at a rate that depends on source declination and hour angle. For the science target, the total integration time will be broken up sinto a dozen or more 5 minute exposures in order to prevent any rotational blurring of the disk image. Each exposure consists of four simultaneous images in adjacent optical bandpasses. The reference star is chosen to match the science target's brightness and color, and for a location relatively nearby on the sky. Rotational blurring is not important for featureless reference stars, allowing for longer integration times. For both stars, initial unsaturated stellar images are taken to acquire the target and center it behind the coronagraphic mask.  

Residual scattered starlight forms a field of speckles with a surface brightness similar to a disk's. Careful data reduction is required to subtract the speckles in order to reveal faint circumstellar sources. Given the relative stability of the wavefront due to the thin atmosphere, especially compared to ground-based telescopes, the DDX speckle field can be largely subtracted out with various calibration techniques. Our baseline procedure involves subtracting each science image by the image of a reference star that is appropriately registered in position and intensity scaled. Because of field rotation, all of the speckle-subtracted science images will then be rotated to a common orientation and combined. Other subtraction techniques are available and can be applied to DDX data, but this method has been chosen as the default as it is robust and has been the most widely used for HST debris disk observations.

The quantitative exploitation of the results starts with direct measurements of disk radius, extent, surface brightness, radial profile, centration relative to the star, and asymmetries in all four imaging bandpasses. The presence of any confusing background sources, such as galaxies present in the HST fields of the AU Mic and HD 107146 debris disks, is defined from their morphology and colors. The next step is to derive a model dust density distribution whose appearance in scattered light best matches the observed images. This will be done with existing codes that treat one or more radial zones with power laws for the surface density profile and a Henyey-Greenstein scattering phase function. 

The best-fit dust spatial distribution found from the modeling will be directly compared to dynamical models for disk structure. Of particular interest are sharp disk edges, isolated azimuthal clumps, a narrow ring morphology, or any ring eccentricity - all direct indications of an adjacent planetary perturber. The images will be compared to dynamical models to infer the mass and location of the planet. These measurements of disk dynamical distrurbances provide the only avenue for measuring the presence and properties of planets in the outer reaches of nearby star systems. 

The total scattered light brightness of the disk, expressed as a fraction of the direct starlight, is then calculated from these models for all four imaging bands. Changes in the scattered light fraction vs. wavelength directly quantify the disk color. The DDX imaging data thus provides two direct constraints on the dust grain size: the degree of forward scattering (which at optical wavelengths increases with grain size) and the disk color (which becomes blue for smaller grains). The scattered light fraction, in combination with the fractional infrared luminosity of the disk measured from the far-infrared spectral energy distribution (SED), allows the dust albedo to be calculated. The derived albedo provides a third constraint on the dust composition, with high values suggesting icy grains and low values indicative of organics that have darkened from radiation exposure.

A crucial follow-on modeling step is to combine the DDX disk model with the far-infrared spectral energy distribution (SED) of the disk measured by Spitzer and Herschel. Such data is available for all DDX targets. Fixing the radial dust distribution to the results of the DDX model fits, the expected far-infrared excess emission from that dust can be calculated from Mie theory for any assumed grain size distribution and composition. Experience in combined model fits of HST and Spitzer data has shown that these models are very sensitive to the minimum grain size, a crucial parameter that establishes the dust dynamical lifetime against removal by radiative forces. The SED fit based on the DDX measurements thus provides a fourth direct constraint on the disk grain properties.

\section{Summary}
 
DDX will address two key elements of the NASA Strategic Plan: to understand the many phenomena and processes associated with planetary system formation and evolution and generate a census of extrasolar planets and measure their properties.  It will do so by placing the Solar System in context by imaging debris disks around nearby stars, potentially revealing the presence of perturbing planets via their influence
on disk structure and will explore the planet formation history of each observed system.  It will determine if the Solar System's two belt architecture is normal and how dust is produced and transported in debris disks.

In addition to the science DDX will carry out, it will also support technology development for future coronagraphs. DDX is a cost-effective means of demonstrating the science potential of the coronagraph instrument, and with actual flight experience, it would raise the technology readiness level and provide heritage for the development of a flight coronagraph for WFIRST. The cost of DDX is a fraction of a probe or flagship space mission and is a cost effective way to reduce technical risk.


\acknowledgments     

This research was carried out at the Jet Propulsion Laboratory, California Institute of Technology, under a contract with the National Aeronautics and Space Administration.



\bibliography{ddx}   

\bibliographystyle{spiebib}   

\end{document}